\documentclass[twocolumn,pra,amssymb,showpacs]{revtex4-1}
\usepackage{graphicx}
\usepackage{amsmath}

\newcommand{\be}{\begin{equation}}
\newcommand{\ee}{\end{equation}}
\newcommand{\ben}{\begin{eqnarray}}
\newcommand{\een}{\end{eqnarray}}

\newcommand{\nd}{\noindent}

\begin{document}

\title{Maximally correlated multipartite quantum states}
\author{J. Batle$^{1} $, M. Casas$^{2}$, A. Plastino$^{2,\,3}$}
\affiliation{ $^1$Departament de F\'{\i}sica, Universitat de les
Illes Balears,
 07122 Palma de Mallorca, Spain  \\\\
$^2$Departament de F\'{\i}sica and IFISC, Universitat de les Illes Balears,
 07122 Palma de Mallorca, Spain \\\\$^3$ Instituto de F\'{\i}sica La Plata--CCT-CONICET, Universidad Nacional de La Plata, C.C.~67, 1900,
La Plata, Argentina}
\date{\today}

\begin{abstract}

\nd We investigate quantum states that posses both maximum
entanglement and maximum discord between the pertinent parties.
Since entanglement (discord) is defined only for bipartite (two
qubit) systems, we shall introduce an appropriate  sum over of all
bi-partitions as the associated  measure. The ensuing  definition
--not new for entanglement-- is thus extended here to quantum
discord. Also, additional dimensions within the parties are
considered ({\it qudits}). We also discuss nonlocality (in the form
of  maximum violation of a Bell inequality)  for all multiqubit
systems. The emergence of more nonlocal states than local ones, all
of them possessing maximum entanglement, will be linked,
surprisingly enough, to whether  quantum mechanics is defined over
the fields of real or complex numbers.


\end{abstract}

\pacs{03.67.-a; 03.67.Mn; 03.65.-w}

\maketitle

\section{Introduction}

\nd Entanglement, non-locality, and quantum discord are foundational
quantum mechanical issues  \cite{Nielsen} and have received renewed
and intense attention in the last years (for a very small sample,
see for instance \cite{Nielsen,Vidal,Latorre,measures} and
references therein). Some important distinctions between the three
concept deserve a few words. Entanglement, non-locality, and quantum
discord are foundational quantum mechanical issues  \cite{Nielsen}
and have received renewed and intense attention in the last years
(for a very small sample, see for instance
\cite{Nielsen,Vidal,Latorre,measures} and references therein). Some
important distinctions between the three concept deserve a few
words.

\begin{itemize}

\item While two-party entanglement is quite well understood,
entanglement in a multi-partite system is an area of great current
interest (see \cite{noltros} and references therein). Since
entanglement is a resource to be taken advantage of, the quest for
maximally entangled states has proceeded at a rapid pace. Thus far,
only few qubit maximally entangled states such as two qubit Bell
states, three qubit Greenberger-Horne-Zeilinger (GHZ) states, and
four qubit states \cite{HS} have been clearly identified. Of course,
it would be of considerable interest to generate such maximally
entangled states as representatives of the ground state of a
physically realizable (spin) model. Why? Because the strength of
correlations in a many- body system is a reflection of the degree of
entanglement (for pure states) \cite{Latorre}. We are here speaking
of quantities such as density, magnetization, etc. Thus,
characterization of multi- particle entanglement and production of
maximal/high multi-qubit entanglement is vital for the mutual
enrichment of i) quantum information physics and ii) many-body
condensed matter.

\item There exist certain tasks, such as device-independent quantum
key distribution \cite{device} and quantum communication complexity
problems \cite{comcomplex}, which can only be carried out provided
the corresponding entangled states exhibit {\it nonlocal
correlations}. Non-locality, as measured by the violation of Bell's
inequalities, describes the part of quantum correlations that cannot
be reproduced by any classical local model. Non-locality plays a key
role in some applications of quantum information theory
\cite{device} and for infinite quantum system \cite{noltros}, in
which  violation of Bell's inequalities can constitute a
complementary resource and pinpoint  the critical points associated
with quantum phase transitions.

\item In turn, quantum discord \cite{6} is defined as the
difference between two expressions of mutual information extended
from the classical to the quantum realm. It aims to capture all the
nonclassical correlations of a quantum system, something that
entanglement fails to do.

\item In recent years, a large number of studies focus attention on
strictly quantal correlations \cite{7,8,9,10,11,12,13,14,15}, and
many different correlation measures have been proposed to detect
them \cite{16,17,18,19,20,21}. It is of the essence to point out
that these correlations, and their measures, exhibit intriguing
peculiarities.  Thus, violation of Bell's inequalities implies
entanglement, but the presence of entanglement does not necessarily
imply Bell-violation \cite{22,23}. In turn, quantum discord exhibits
a distinctly different behavior from that of entanglement \cite{25}.
Remarkably enough, quantum discord has proved itself a useful
quantity in a number of applications, such as speed-up in quantum
computation \cite{28}.
\end{itemize}

\nd For our present purposes we specially note that quantum
teleportation --one of the most intriguing features of
entanglement-- has been proposed theoretically using protocols based
on genuine {\it multipartite} entanglement, and experimentally
realized in four and five qubits systems \cite{Yeo,Zhao}. States
with genuine multipartite entanglement are defined as having  strict
maximal entanglement in all bipartitions of the system. The question
of maximal entangled states and their definition was first addressed
by Gisin \cite{Gisinmixed}, where the existence of only a few of
them was first pointed out. Along  lines recently pursued  by Helwig
et al. \cite{Helwig}, we shall focus attention here on states that
may become maximally entangled for any number of parties {\it by
changing the qudit dimension}. One clear motivation for studying
these states is that their maximal entanglement content makes them
natural candidates for implementing original multipartite
communication protocols.

\nd The concept of quantum correlation for multipartite systems is
intimately related to the mathematical structure of quantum
mechanics, as a direct consequence of the linear character of tensor
product Hilbert spaces. A question that will seen important below
revolves around the fact that the Hilbert space of a system may use
the field of real or of complex numbers. We will show instances in
which a description provided by real coefficients is both useful and
necessary.

\nd  \fbox{\parbox{0.97\linewidth}{ It is the aim of the present
work to  regard multipartite quantum correlations from three
standpoints: those of  quantum entanglement, quantum discord, and
non-locality.}}\newline \newline \nd    In the Appendix  we review
all measures employed for describing quantum correlations. Some
readers may want to peruse it before proceeding  with the paper.
Equations in the Appendix are numbered with a capital A preceding
the corresponding number and are referenced throughout  the text.
In Section II we study maximally entangled states for N=3 till 8
qubits, emphasizing the role played, in each case, by
non-locality. Section III is devoted to obtaining maximally
entangled states for systems of qudits. States are compared with
their corresponding qubit counterparts. Section IV makes use of
quantum discord in order to find multipartite qubit system
maximizing it. Section V compares all correlation measures by
individually studying how they evolve as the number of parties
increases, or globally correlating their absolute values for
different maximal states with fixed number of parties. Finally,
some conclusions are drawn in Section VI.

\section{Maximally entangled states of multiqubit systems. The role of nonlocality}

 \nd When dealing with states maximizing entanglement in multiple qubits, it will prove
  convenient to employ the maximally correlated basis
that is defined for every $2^N$-Hilbert space. Usually, the
preferred basis for two qubit states is the so called computational
one $\{ |00\rangle ,|01\rangle,|10\rangle,|11\rangle \}$. The  Bell
basis of maximally correlated states for N=2 is given by

\begin{equation} \label{bellbasis}
|\Phi^{\pm} \rangle= \frac{(|00\rangle \pm e^{i\theta}|11\rangle)}{\sqrt{2}}, |\Psi^{\pm} \rangle= \frac{(|01\rangle \pm e^{i\theta}|10\rangle)}{\sqrt{2}}.
\end{equation}

\noindent For simplicity, and without loss of generality, we shall
take real coefficients ($\theta=0$ in (\ref{bellbasis})). These
states maximally violate the CHSH Bell inequality. Needless to say,
only in the bipartite case entanglement implies nonlocality and vice
versa ({\it Gisin's theorem} \cite{GisinTheorem}). Let us recall
that the family of pure states of three qubits $|\Psi_{j,N=3}^{ \pm
} \rangle =(|j\rangle \pm |2^{N=3}-1-j\rangle)/\sqrt{2}$ forms a
basis, the so called GHZ basis or Mermin basis, and that these
states maximally violate the Mermin inequality. In the general case,
the most convenient basis for dealing with maximally correlated
multipartite states of N qubits will be given by

\begin{equation}
|\Psi_{j,N}^{ \pm } \rangle =(|j\rangle \pm |2^{N}-1-j\rangle)/\sqrt{2}.
\end{equation}

\noindent Notice that these states maximally violate the generalized
MABK Bell inequality.

 \nd For N=3 qubits, the maximally entangled state is given by
 the usual GHZ state $|GHZ\rangle=(|000\rangle+|111\rangle)/\sqrt{2}$.
This instance will be the only one for which maximum entanglement
and maximum non-locality are provided by the same state. For N=4
qubits, the state discovered by Higuchi and Sudbery \cite{HS},
given by

\be \label{HS} |\Phi_4 \rangle \, = \, \frac{1}{\sqrt{3}}
\Bigl[ |\Psi_{3,4}^{+} \rangle + \omega |\Psi_{6,4}^{+} \rangle+
\omega^2 |\Psi_{5,4}^{+} \rangle \Bigr], \ee

\noindent with $\omega= -\frac{1}{2}+\frac{\sqrt{3}}{2}i$, is the
one which possesses maximum entanglement ($9.3773\ln2$) as defined
by the measure of Eq. (\ref{Smesure}). For this entanglement
measure to be optimal, the corresponding state must possess
complex expansion-coefficients. In addition, $QD \ne 0$ for state
(\ref{HS}). When pursuing maximum entanglement for real states
{\it only}, we encounter the following state

\begin{equation} \label{estat4real}
 \begin{array}{rl}
 |\Phi_4^{R} \rangle \, = \, \frac{1}{2\sqrt{2}} \, \Big[ \, & |\Psi_{0,4}^{-} \rangle -|\Psi_{3,4}^{+} \rangle- |\Psi_{4,4}^{+} \rangle   \\
&+ \sqrt{2}|\Psi_{5,4}^{+} \rangle +\sqrt{2} |\Psi_{6,4}^{-} \rangle+ |\Psi_{7,4}^{-} \rangle \big) \, \Big].
\end{array}
\end{equation}

\noindent This state--which has null discord-- is highly entangled,
but does not reach the maximal value. Instead, it stays very close
to it ($9.2017\ln2$). Specifically, for the state (\ref{estat4real})
 all one qubit reduced matrices are maximally mixed, as in the
case of the state (\ref{HS}), and two of the six associated
density matrices for two qubits are maximally mixed. In the light
of these results, we should require some figure of merit to
somehow validate one of these two states or both of them. Let us
suppose that we study the concomitant MABK Bell inequality maximal
violation given in the Appendix by Eq. (\ref{MABK_Nmax}). In the
complex case of four qubits, $MABK_4^{\max} = 2.17732$, a $38\%$
of the maximum possible violation. On the other hand, our real
state of four qubits has $MABK_4^{\max} = \sqrt{6}$, only a $43\%$
of the maximum value. This is not a big surprise, for the more
linear combinations of maximally correlated states we have, the
less non-locality is attained \cite{noltros}. In our case,
however, $|\Phi_4^{R} \rangle$ is more nonlocal than $|\Phi_4
\rangle$, although less entangled.

 \nd In this case, the state possessing maximum correlations,
 either in the form of entanglement or of non-locality, is not the one with complex coefficients. Therefore, it makes a
great difference which  field of numbers we use for building up
quantum states.

\nd For N=5 qubits, Brown et al. \cite{BSSB05} proposed a state
which is shown to possess the maximum entanglement given in he
Appendix by (\ref{Smesure}). This state of five qubits is of the
form

\be
|\Phi_5 \rangle \, = \, \frac{1}{2} \Bigl[ |100\rangle
|\Phi_{-}\rangle +|010\rangle |\Psi_{-}\rangle +
|100\rangle|\Phi_{+}\rangle + |111\rangle|\Psi_{+}\rangle\Bigr],
 \ee

\noindent with a maximum entanglement of $25\ln2$. It is apparent
that the previous state does not contain correlations going beyond
those for  the bipartite case. Certainly, the concomitant
non-locality measure is 2.1361, only $27\%$ of the corresponding
maximum possible value.

 \nd For N=6 qubits, we discovered in Ref. \cite{noltrostots} a state with a maximum entanglement
given by Eq. (\ref{Smesure}). This is a state for which all
bipartitions are maximally mixed. The aforementioned state has the
form

 \begin{equation} \label{estat6}
 \begin{array}{rl}
 |\Phi_6 \rangle \, = &\, \frac{1}{4} \, \Big[ \, |\Psi_{0,6}^{+} \rangle +|\Psi_{3,6}^{+} \rangle+ |\Psi_{5,6}^{+} \rangle  + |\Psi_{6,6}^{+} \rangle \\
&+ |\Psi_{9,6}^{+} \rangle+ |\Psi_{15,6}^{+} \rangle +|\Psi_{17,6}^{+} \rangle + |\Psi_{18,6}^{+} \rangle  \\
&+ |\Psi_{24,6}^{+} \rangle + |\Psi_{29,6}^{+} \rangle - \big( \, |\Psi_{10,6}^{+} \rangle + |\Psi_{12,6}^{+} \rangle \\
&+ |\Psi_{20,6}^{+} \rangle + |\Psi_{23,6}^{+} \rangle +|\Psi_{27,6}^{+} \rangle + |\Psi_{30,6}^{+} \rangle \big) \, \Big].
\end{array}
\end{equation}

\noindent Remarkably enough, Tapiador et al. \cite{Tapiador} were
able to algebraically simplify this state into a form that greatly
clarifies how correlations are distributed among subsystems. This
state $|\Psi_6\rangle$ reads

\begin{equation} \label{estat6anglesos}
 \frac{1}{2}(
|\Psi_{0,4}^{+} \rangle |\Psi^{-}\rangle + |\Psi_{3,4}^{+} \rangle |\Psi^{+}\rangle +
|\Psi_{6,4}^{+} \rangle |\Phi^{-}\rangle + |\Psi_{5,4}^{+} \rangle |\Phi^{+}\rangle).
\end{equation}

\noindent It is plain from (\ref{estat6anglesos}) that all
correlations existing in this maximally entangled state are
encoded via maximally correlated 4-party and 2-party subsystems,
which entails interesting implications regarding non-locality.
Indeed, the maximal violation of the corresponding MABK Bell
inequality for 6 qubits is only 2 ($18\%$ of the maximum value).
In other words, state (\ref{estat6anglesos}) may have maximum
entanglement, but does not violate the corresponding MABK
inequality. For this state  all concomitant reduced density
matrices of one, two and three qubits are completely mixed, yet
the state  is not nonlocal. $|\Psi_6\rangle$ constitutes an
example of a state which has found applications in quantum
information processing \cite{t1,t3}, but since it has negligible
non-locality, it may not be suitable for other tasks
\cite{device,comcomplex}.


\subsection{Special cases: N=7 and N=8 qubits}

\nd These two particular instances deserve  special consideration
for --as far as we know-- they have not yet been investigated in
detailed fashion.

In the case of N=7 qubits, since the state obtained in
\cite{noltrostots} is one with no easily discernible algebraic
structure, one wonders whether a better form for that state might
exist, either for real or complex state-functions. A more detailed
study is  carried out here. A  conjecture made in
\cite{noltrostots} establishes that there is no pure state of
seven qubits with  marginal density matrices for subsystems of
one, two, or three qubits that are (all of them) \emph{completely}
mixed. Contrarily to the case of the seven-qubit state reported in
\cite{noltrostots}, we  encounter other states that --although
lacking a simple structure-- all possess totally mixed 1- and
2-qubit marginal density matrices. Several highly entangled states
of 7 qubits have been obtained \cite{Tapiador} without reaching an
upper bound.

\nd The aforementioned state of N=7 qubits  possesses a maximum
entanglement of 105.7882. Certainly, its  reduced states of two
qubits are not (all) mixed, but in practice, they can be
considered as such. This is so because, for reduced states of two
qubits, 6 of them possess entanglement $0.99205\ln2$ and the
remaining 15, $0.99760\ln2$.  This ``imperfect'' mixture surely is
immaterial for practical purposes. As far as 3-qubit marginal
density matrices are concerned, 15 contain an entanglement of
$0.97913\ln8$ and the remaining 20, $0.99541\ln8$. Although some
more compact algebraic form for the seven-qubit maximally
entangled state could in principle be found, that is not actually
the case. Such putative  state should necessarily  be complex
--not real-- to attain maximum entanglement, which is paramount
importance for our present discussion.


In the light of these results, we may wander if it makes {\it physical} sense to have states
 that are not completely mixed. Should  one expect
a maximally entangled state to be endowed with  some sort of
hierarchy in which increasing entanglement would be accompanied by
 maximal mixedness at every one-, two-, three-stages involving
marginal states? Also, is it mandatory to resort to states with
complex coefficients in order to have maximum multipartite
entanglement? These are  precisely the questions we wish to answer
 in the present work.

To this end we have explored the whole space of real and complex
states via a simulated annealing method, and have found similar
results for both real and complex states. We have been able to
detect states with totally mixed 1- and 2-qubit marginal density
matrices, while increasing mixedness for 3 qubits. In the case of
N=7 real states, we encountered a maximum entanglement of
105.1151, greater than that recorded all previous findings. On the
other hand, the complex expansion-coefficients instance displays a
slightly greater value of entanglement than the real one, namely,
105.1441. While the former situation accrues  nearly 31/35
completely mixed three qubit reduced density matrices, in the
later one that figure is increased up to 33/35. In Fig. 1, the
coefficients of the real state are depicted. Had we let the
optimization procedure evolve freely {\it in the complex case}, we
would have reached the same result as in \cite{noltrostots}. In
other words, it is possible to obtain highly entangled states,
either real or complex, with totally mixed 1- and 2-qubit reduced
states. To the best of our knowledge, no other states had been
previously provided with this maximal entanglement values.

\begin{figure}[htbp]
\begin{center}
\includegraphics[width=8.6cm]{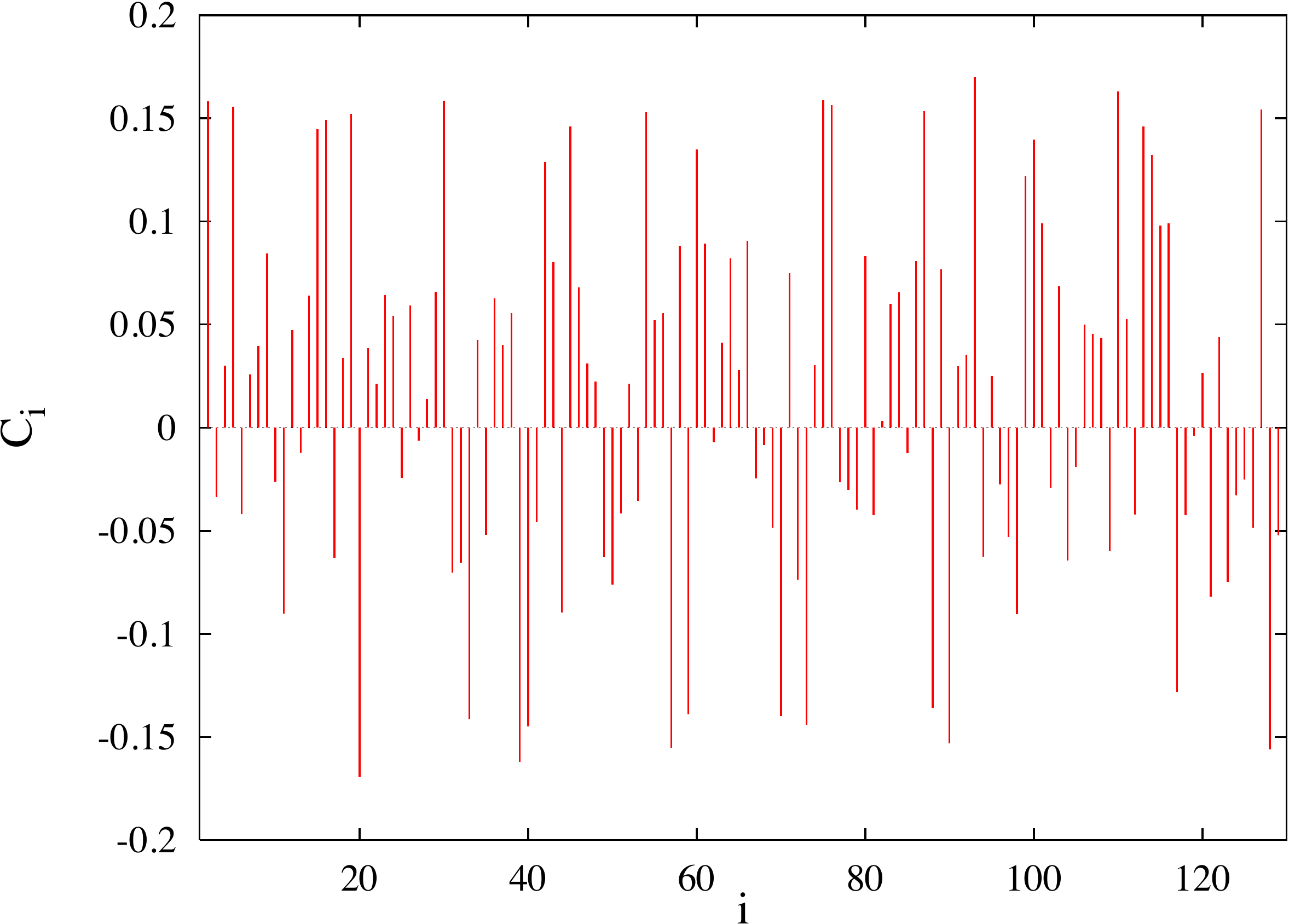}
\caption{(Color online) Coefficients of the real state of $N=7$
qubits that maximizes total entanglement, with fully 1- and
2-qubit reduced states that are maximally mixed . See text for
details.} \label{fig1}
\end{center}
\end{figure}

\nd On the whole, although we have not been able to provide a
conclusive answer regarding entanglement for the N=7 qubit case,
we have shed light on the characteristics of these states by
introducing a new one that possesses  maximally mixed reduced two
qubit states.
 As we have seen,
real or complex states behave differently, a situation that might
become immaterial, provided real states suffice to fulfill
eventual applications. As far as non-locality is concerned, all
$N=7$ qubit instances do not violate the corresponding MABK
inequality. Again, as in the $N=6$ case, we find completely
opposite behaviors for entanglement vis-a-vis non-locality.

\nd Now, the last case of interest, which has been quite elusive,
is the one corresponding to N=8 qubits. It is undeniable that as
we increase more and more the dimension of the concomitant Hilbert
space, the problem of obtaining an optimum state with maximum
generalized entanglement becomes less and less tractable, either
analytically or numerically.

\nd In any case,  we have carried out extensive computations in
order to get a suitable result. By performing a simulated
annealing optimization procedure, we have reached a maximum for
the entanglement and, in turn, obtained a multipartite state of
eight qubits that appears to be an optimal one. We  mention that
in order to tackle the problem, we have only pursued real states,
that is, configurations of states with real coefficients. This
fact may limit the validity of our conclusions, but we are
confident that, in any case, we have gotten  a valid result. In
addition, we have limited the expansion-coefficients to be all
equal in modulus, thus acquiring a {\it simple} algebraic
structure. Needless to say, these simplifications have led to
greater values of entanglement than arbitrary explorations with no
constraints.

\nd Our special  state of eight qubits $|\Phi_8 \rangle$ reads

 \begin{equation} \label{estat8}
 \begin{array}{rl}
 \frac{1}{4\sqrt{2}}  \Big[
& - |\Psi_{0,8}^{-} \rangle -|\Psi_{3,8}^{-} \rangle+ |\Psi_{5,8}^{-} \rangle  - |\Psi_{6,8}^{-} \rangle - |\Psi_{25,8}^{+} \rangle \\
& + |\Psi_{26,8}^{+} \rangle +|\Psi_{28,8}^{+} \rangle+ |\Psi_{31,8}^{+} \rangle  + |\Psi_{33,8}^{-} \rangle - |\Psi_{34,8}^{-} \rangle \\
& + |\Psi_{36,8}^{-} \rangle + |\Psi_{39,8}^{-} \rangle+ |\Psi_{56,8}^{+} \rangle  + |\Psi_{59,8}^{+} \rangle + |\Psi_{61,8}^{+} \rangle \\
& - |\Psi_{62,8}^{+} \rangle +|\Psi_{73,8}^{-} \rangle+ |\Psi_{74,8}^{-} \rangle  + |\Psi_{76,8}^{-} \rangle - |\Psi_{79,8}^{-} \rangle \\
& + |\Psi_{80,8}^{+} \rangle -|\Psi_{83,8}^{+} \rangle+ |\Psi_{85,8}^{+} \rangle  + |\Psi_{86,8}^{+} \rangle - |\Psi_{104,8}^{-} \rangle \\
& + |\Psi_{107,8}^{-} \rangle +|\Psi_{109,8}^{-} \rangle+ |\Psi_{110,8}^{-} \rangle  - |\Psi_{113,8}^{+} \rangle \\
& - |\Psi_{114,8}^{+} \rangle + |\Psi_{116,8}^{+} \rangle -|\Psi_{119,8}^{+} \rangle \big) \, \Big].
\end{array}
\end{equation}

\noindent It exhibits interesting features. The amount of total
entanglement is $362\ln2$. All reduced states of one, two, and
three qubits are maximally mixed. This fact implies that this
state is perfectly suitable for i) performing teleportation
protocols involving up to three qubits within eight parties, as
well as ii) for many other applications. However, as expected, all
four qubit reduced states are not completely mixed. The work by
Gisin and Bechmann-Pasquinucci \cite{Gisinmixed} already pointed
out that states of eight qubits cannot have all subsystems
completely mixed. However, in our case, we get close enough to
such desideratum.

\noindent Out of the $\binom {8} {4}=70$ possible reduced states of four qubits,

\begin{itemize}
\item 56 matrices are completely mixed ($\ln16$)
\item 6 of them are nearly diagonal matrices with four equal
eigenvalues ($\frac{1}{2}\ln16$). The corresponding partitions are
$\{1,2,3,6\},\{1,2,4,5\},\{1,2,7,8\},\linebreak \{3,4,5,6\}, \{3,6,7,8\},\{4,5,7,8\}$.
\item The 8 remaining matrices are almost diagonal,
possessing eight equal eigenvalues ($\frac{3}{4}\ln16$). The concomitant partitions are
$\{1,3,4,7\},\{1,3,5,8\},\{1,4,6,8\},\{1,5,6,7\},\linebreak \{2,3,4,8\},\{2,3,5,7\},\{2,4,6,7\},\{2,5,6,8\}$.
\end{itemize}

\nd As far as non-locality is concerned, state (\ref{estat8}) does
not violate the corresponding MABK Bell inequality. In the light
of the previous results, we may formulate the following
conjecture.

\noindent {\bf Conjecture 1.} {\it No maximally entangled state
greater that 5 qubits violates the MABK Bell inequalities}.

\noindent We shall see in forthcoming sections that non-locality
for maximally entangled states decreases exponentially with the
number of qubits.

\nd Summing up, 1) we confirm that no states of N=8 qubit exist
with all their marginal density matrices being maximally mixed and
2) We provided an example of real state that might be  tailored
for quantum communication protocols or teleportation.

\section{Maximally entangled states of higher dimensions}

\nd Maximum {\it qudit} entanglement  constitutes an extension of
a previous study for qubits to states living in a $D^N$-Hilbert
space, where $D$ stands for the dimension of each party.  Helwig
et al. \cite{Helwig} addressed the interesting and nontrivial
problem of finding the conditions for the existence of states that
maximize the entanglement between all bipartitions. These states
are of the type $|\Psi_{N,D}\rangle=\sum_{i=0}^{D^N-1} c_i
|i\rangle$.   The equivalence between several pure state quantum
secret sharing schemes and states with maximum multipartite
$(N,D)-$entanglement with an even number of parties is proven in
\cite{Helwig}, an equivalence which indirectly implies the
existence of these maximally entangled states for an arbitrary
number of parties, based on known results about the existence of
quantum secret sharing schemes.

\nd We will try to ascertain just what sort of states can host a
maximum amount of entanglement between their parties. Since no
general procedure has yet been  provided  for studying what
conditions these maximally entangled qudit states should fulfill,
we shall resort to numerical explorations that will hopefully shed
some insight into qudits-systems that might not apply for
qubits-ones. As dimensional-examples  we have considered the cases
${\cal H}_3^{\otimes 3}$,  ${\cal H}_4^{\otimes 3}$, ${\cal
H}_3^{\otimes 4}$ and ${\cal H}_3^{\otimes 5}$.

\nd The case of three qutrits (${\cal H}_3^{\otimes 3}$) is really
exceptional, as we shall see. The Hilbert space ${\cal H}_3$ is
spanned by the basis $\{ |0\rangle,|1\rangle,|2\rangle \}$. The
state $|\Psi_{N=3,D=3}\rangle$

\begin{equation} \label{estatD3}
 \frac{1}{\sqrt{6}} \Big[ \,  |000 \rangle -|011 \rangle -|112 \rangle  + |120 \rangle - |202 \rangle+ |221 \rangle  \, \Big]
\end{equation}

\noindent possesses maximum entanglement $3\ln3$ and from
inspection, it is biseparable. The situation becomes more involved
when one notices that the state

\begin{equation} \label{estatD3p}
|\Psi_{N=3,D=3}^{'}\rangle =  \frac{1}{\sqrt{3}} \Big[ \,
|000 \rangle + |111 \rangle  + |222 \rangle  \Big],
\end{equation}

\noindent which is clearly nonlocal, also reaches the maximum
value $3\ln3$. How is it possible that two  different states (one
being separable, and the other  non-separable) may attain the same
entanglement-based correlations? Should we use (\ref{estatD3p})
and regard state  (\ref{estatD3}) as an anomaly, or should we
revisit instead the definition of maximum entanglement?

\noindent Increasing individual party dimensions by one unit, we
get a state living in the Hilbert space ${\cal H}_4^{\otimes 3}$
(spanned by $\{ |0\rangle,|1\rangle,|2\rangle ,|3\rangle\}$). The
maximally entangled state for $N=3,D=4$ $|\Psi_{3,4} \rangle $
reads

\begin{equation} \label{estat43}
 \frac{1}{2\sqrt{2}} \Big[ \,  \sqrt{2} |002 \rangle + \sqrt{2}|310 \rangle -|121 \rangle
 + |123 \rangle + |231 \rangle+ |233 \rangle  \, \Big].
\end{equation}

\noindent This dimension does not pose the puzzle we found
previously for three qutrits, in the sense that it is maximally
entangled ($3\ln4$) as opposed to

\begin{equation} \label{estat43p}
|\Psi_{3,4}^{'} \rangle= \frac{1}{2} \Big[ \,  |000 \rangle + |111 \rangle + |222 \rangle  + |333 \rangle  \, \Big],
\end{equation}

\noindent which has entanglement $\frac{5}{4}\ln4$.  For
(\ref{estat43}), the first two elements are inseparable, while the
remaining ones are biseparable. Thus, the role of quantum
correlations is not as dominant  as above. To be more rigorous, we
should develop some tight Bell inequality and resort to the
concomitant maximum violation. Unfortunately, no such Bell
inequalities have been encountered so far.

\nd The case of four qutrits (${\cal H}_3^{\otimes 4}$) is the
natural extension of the four qubit state. The state we obtain is
of the form

 \begin{equation} \label{estat34}
 \begin{array}{rl}
 |\Psi_{N=4,D=3}\rangle = \frac{1}{6}  \big(
&  |0000 \rangle + |1000 \rangle  + |0021 \rangle +|1021 \rangle +\\
&  |0100 \rangle - |0110 \rangle  - |0121 \rangle -|0122 \rangle +\\
&  |0201 \rangle - |0202 \rangle  + |0211 \rangle +|0220 \rangle \\
&  -|1100 \rangle - |1110 \rangle  + |1121 \rangle -|1122 \rangle \\
&  -|1201 \rangle - |1202 \rangle  + |1211 \rangle -|1220 \rangle \,\,\big) \\
  +\frac{\sqrt{2}}{6}  \big(
 &   -|0012 \rangle + |1012 \rangle  - |2001 \rangle +|2020 \rangle \\
&  -|2102 \rangle - |2111 \rangle  - |2210 \rangle +|2222 \rangle \,\,\big).
\end{array}
\end{equation}

\noindent A careful analysis shows that  the above state can be
written as a combination of tensor products, where only bipartite
correlations appear. These loose correlations in the four-party
case should be  contrasted with the corresponding ones for four
qubit states, where maximum entanglement is reached for i) a state
with complex expansion-coefficients (linear combination of three
maximally correlated states) and ii) real ones (state with high,
but not maximum entanglement embedded into a linear combination of
six maximally correlated states). The differences between the
cases of maximum entanglement between $(N,D=2)$ and arbitrary
$(N,D)$ have to be taken into account in order to shed light on
the  problem of quantifying and characterizing entanglement for
multipartite systems.

\nd The case of five qutrits (${\cal H}_3^{\otimes 5}$) has turned
out to be elusive. We cannot provide a simple expression for the
concomitant state, only an approximate one. However, numerical
evidence shows, in a 'real' quantum treatment, it is most likely
that a maximum entanglement ($25\ln3$) might be reached, that is,
all reduced states of one and two qubits are likely to be
maximally mixed. However,  this is just an approximate result. If
it were confirmed, this would entail that in greater dimensions,
real expansion-coefficients suffice to describe all sorts of
states with maximum correlations. This fact has been confirmed for
the case $(N=3,D=5)$ ($S=3\ln5$) as well, but the concomitant
state can not be casted in  simple fashion. In view of the
previous results, we formulate a second conjecture.

\noindent {\bf Conjecture 2.} {\it Maximally entangled states of
multiqudit systems $(N,D)$ only require real
expansion-coefficients. Also, all their reduced density matrices
are completely mixed}.

\noindent For multiqubit states, in only a few cases  all reduced
density matrices are proportional to the normalized identity (that
is, maximum entanglement), whereas in the case of qudits the
instances encountered do not display this behavior.  Furthermore,
 it is  important to use complex expansion-coefficients for qubits to get
maximum entanglement, but this constraint disappears in higher
dimensions.  These are issues  of great interest that find at least
partial elucidation in the present work.

\section{Multiqubit states with maximal discord}

\noindent In the case of entanglement, the most straightforward way
of tackling quantum correlations in multipartite states is to
introduce partitions into the system. This is somehow inevitable, as
 no definitive entanglement measure or criterion is yet available
for characterizing true multipartite quantum correlations of this
kind. Since $QD$ is defined only between pairs of qubits, it is
quite natural to extend the same tools used for multipartite
entanglement to the case of quantum discord. If not otherwise
stated, all states are given in the computational basis
$\{|000..00\rangle,|000..01\rangle,|000..10\rangle,...,|111..10\rangle,|111..11\rangle
\}$.

\subsection{Geometric discord}

\noindent The GQD, which is a measure introduced so as to grasp all
 properties of the usual discord measure. This is a computable quantity
that detects (and quantifies) true discord. For simplicity, here we
shall compute the maximum GQD just for three qubits.

 \noindent   The case of N=3 qubits is   special,  because the
result we obtain is a rather simple one, but the corresponding state
is instead of an involved nature. When approaching the problem of
finding a state $\rho=|\phi\rangle \langle \phi|$ of three qubits
where $GQD(\rho_{12})+GQD(\rho_{13})+GQD(\rho_{23})$ is maximum, by
definition (Cf. (\ref{Dfinal})), one assumes that the outcome for
the unknown expansion-coefficients is not going to a simple one.
Indeed, by the nature of this measure, we obtain via simulated
annealing what appear to be random coefficients for the state
$\rho=|\phi\rangle \langle \phi|$. This fact should not surprise
anyone since the mathematical restrictions that are imposed when
optimizing the sum of the GQD content for all pairs are highly
nonlinear. Table \ref{1} lists the real and complex coefficients
--in the computational basis-- that yield the same maximum GQD.

\begin{table}[h]
\begin{center}
\begin{tabular}{|c||c|c|}
  \hline
  $Coefficient$ & $Real \, state$ & $Complex \, state$ \\
  \hline
  $c_0$ & 0.435569236 &  (0.0546370323,-0.100659299)  \\
  $c_1$ & -0.186434446 &  (0.0134949587,-0.188315179)  \\
  $c_2$ & 0.151369915 &  (0.465863387,0.0484337848)  \\
  $c_3$ & -0.0680793253 &  (0.242430776,-0.0818648344) \\
  $c_4$ & -0.177771681 &  (0.136790716,-0.217618993)  \\
  $c_5$ & -0.676301307 &  (-0.500174131,0.0275624382)  \\
  $c_6$ & -0.505730715 & (-0.442832499,0.0730590501) \\
  $c_7$ & 0.0567821074 &  (0.0598215336,-0.379958264)  \\
   \hline
\end{tabular}
\end{center}
\caption{Expansion-coefficients for the states of three qubits that
maximize the total GQD between pairs. Columns refers to real and
complex coefficients. See text for details.} \label{1}
\end{table}

\noindent The concomitant maximum GQD is equal to $5/8$,  but {\it
individual} discords are different! That is, $GQD(\rho_{12})=1/8$,
$GQD(\rho_{13})=1/4$, and $GQD(\rho_{23})=1/4$. Apparently, there is
no a priori reason for this to be the case. This asymmetry between
pairs constitutes the first discord feature that is different from
those pertaining to the generalized entanglement measure in
multipartite systems. Strictly speaking, states with maximal
entanglement do have reduced states with different entanglement
values, but this occurs only for large number of reduced-state,
qubit bipartitions. This sort of 'GQD-asymmetry' occurs already in
the simplest possible case of $N=3$ qubits and poses a conundrum.

\subsection{Discord measure}

\noindent Although there is an alternative measure for computing the
discord content of a given state, we  prefer the usual quantum
discord definition QD (\ref{QDmeasure}). We stress here that the
computation of the maximum QD is a two-fold optimization procedure:
first, one must find, given an arbitrary state, the minimum QDs for
all pairs $\rho_{ij}$, and then, survey all states until the maximum
QD is attained.

\noindent It would appear that since N=3 qubits is a low dimensional
system, the state maximizing the concomitant QD should be given by
an algebraically simple expression. As in the case of the previous
GQD measure, this is not the case. \noindent The N=3 qubits case
leads the result $QD=1.662026$, given by the state

 \begin{equation} \label{estatQD3}
 \begin{array}{rl}
 |\Psi^{QD}_3\rangle\,= &\, 0.52895 |000 \rangle +0.19492 |001 \rangle +0.25144 |010 \rangle \\
 & -0.48224 |011 \rangle+0.38250 |100 \rangle -0.39801 |101 \rangle \\
 & -0.20213 |110 \rangle +0.20209 |111 \rangle.
\end{array}
\end{equation}

\noindent The same asymmetry already found  in the QGD case occurs
also here. Also, real and complex states provide the same result.
The non-locality of state $\ref{estatQD3}$ is given by
$MABK_3^{\max} = 3.0430$, a $76\%$ of the maximum possible
violation, with an entanglement $=2.7394\ln2$ ($91\%$, very close to
the maximal one). Thus, a state with maximum $QD$ has a relatively
high amount for the several  quantum correlations surveyed in this
work.

\begin{figure}[htbp]
\begin{center}
\includegraphics[width=8.6cm]{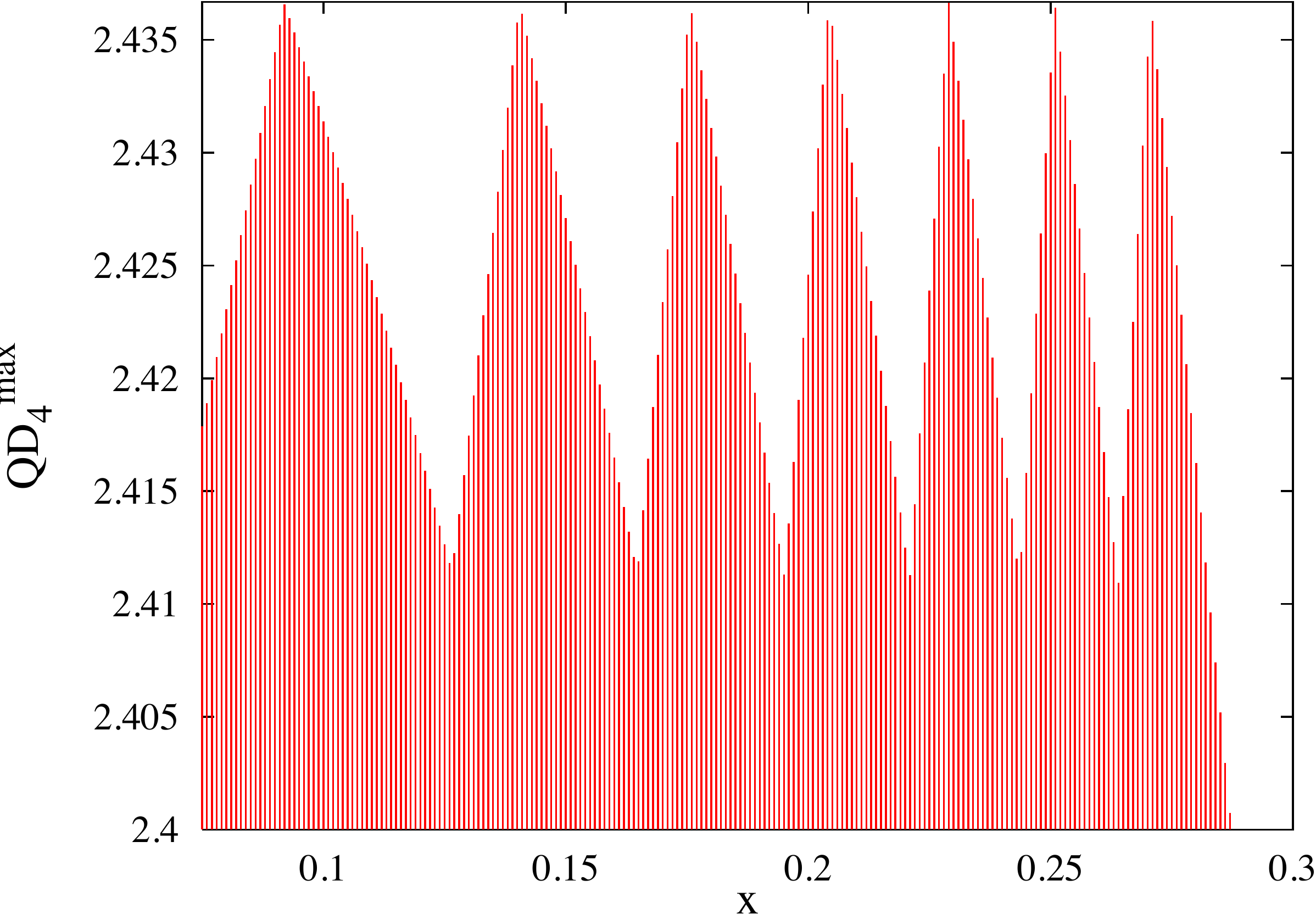}
\caption{(Color online) $QD$ for the family of states (\ref{estatQD40}) of $N=4$ qubits. Seven clear $x-$values reach the maximum value.
The $x=0$ case is given by (\ref{estatQD4}). See text for details.}
\label{fig2}
\end{center}
\end{figure}

\noindent The N=4 case leads the result $QD=2.436681$, with
partitions $\{1,2\},\{3,4\}$ having QD=1/3, and QD=0.442503 for the
remaining pairs. This symmetry also occurs in the case of the
computation of the GQD,  equal to 1.045667. We find the state
$|\Psi^{QD}_4\rangle$ to be of the type

\begin{equation} \label{estatQD40}
 x\big(|\Psi_{2,4}^{-} \rangle
 +\frac{1}{2}|\Psi_{4,4}^{-} \rangle -\frac{1}{2}|\Psi_{7,4}^{-} \rangle \big)+z|\Psi_{3,4}^{+} \rangle+y|\Psi_{5,4}^{+} \rangle +y|\Psi_{6,4}^{+} \rangle.
\end{equation}

\noindent For this family of states, its QD is plotted in Fig. 2 as
a function of $x$. Except for $x=0$, there exists seven additional
states that reach the maximum possible QD-value. The one we consider
here is given in (\ref{estatQD40}) with $x=0$, that is,

\begin{equation} \label{estatQD4}
 |\Psi^{QD}_4\rangle \,=\, \frac{1}{\sqrt{3}} |\Psi_{3,4}^{+} \rangle+\frac{1}{2\sqrt{3}}|\Psi_{5,4}^{+} \rangle +\frac{1}{2\sqrt{3}}|\Psi_{6,4}^{+} \rangle.
\end{equation}

\noindent Notice that this state has all their two qubit reduced
states of the 'X-form'. Accordingly,  their QD is given in
analytical fashion \cite{Xstates}. Remark that  the state that we
have found to attain maximal QD does have the same maximally
correlated sub-states as  the maximally entangled state of four
qubits (\ref{HS})! Needless to say, real and complex states reach
the same maximum QD-value. Regarding non-locality,
$MABK(|\Psi^{QD}_4\rangle)=\frac{8}{3}\sqrt{2}$ ($67\%$).

\noindent In the case of N=5 qubits, we have not reached a ``nice''
algebraic form for the 'maximal' state. All contributions in the
computational basis possess a nonzero weight,
 and the overall state does not seem to exhibit any symmetry.
 However, within the limits of numerical accuracy,
 all $5(5-1)/2=10$ pairs of qubits seem to have the same amount of QD,
 the total one being 3.642445. This fact implies that
 there ought to be a  simpler form for the aforementioned state. The expansion-coefficients are shown in Fig. 3.
 The non-locality of the state $|\Psi^{QD}_5\rangle$ is $\approx 5.0233$ ($63\%$).

\noindent The case of N=6 qubits, on the other hand, does exhibit a
definite symmetry. All reduced pairs have the same QD-value
(0.350977) and the total QD is 5.264662. As in the case of N=4
qubits, all pairs are of the X-form and, thus, analytically
computable. The final (real) state is of the form

 \begin{equation} \label{estatQD6}
 \begin{array}{rl}
|\Psi^{QD}_6\rangle\,= &\, \frac{1}{4\sqrt{5}}
 \Big[  |\Psi_{0,6}^{+} \rangle +|\Psi_{3,6}^{+} \rangle+ |\Psi_{5,6}^{+} \rangle  + |\Psi_{6,6}^{+} \rangle \\
& + |\Psi_{9,6}^{+} \rangle +|\Psi_{10,6}^{+} \rangle+ |\Psi_{12,6}^{+} \rangle  + |\Psi_{15,6}^{+} \rangle  \\
& + |\Psi_{17,6}^{+} \rangle + |\Psi_{18,6}^{+} \rangle+ |\Psi_{20,6}^{+} \rangle  + |\Psi_{23,6}^{+} \rangle  \\
& + |\Psi_{24,6}^{+} \rangle +|\Psi_{27,6}^{+} \rangle+ |\Psi_{30,6}^{+} \rangle  + 5|\Psi_{29,6}^{+} \rangle  \, \Big].
\end{array}
\end{equation}

\noindent Notice again the tendency towards a large number of linear
combinations of maximally correlated states. Specifically, the state
$|\Psi_{29,6}^{+} \rangle$ is particularly relevant. Again, complex
and real coefficients lead to the same QD-value.

\begin{figure}[htbp]
\begin{center}
\includegraphics[width=8.6cm]{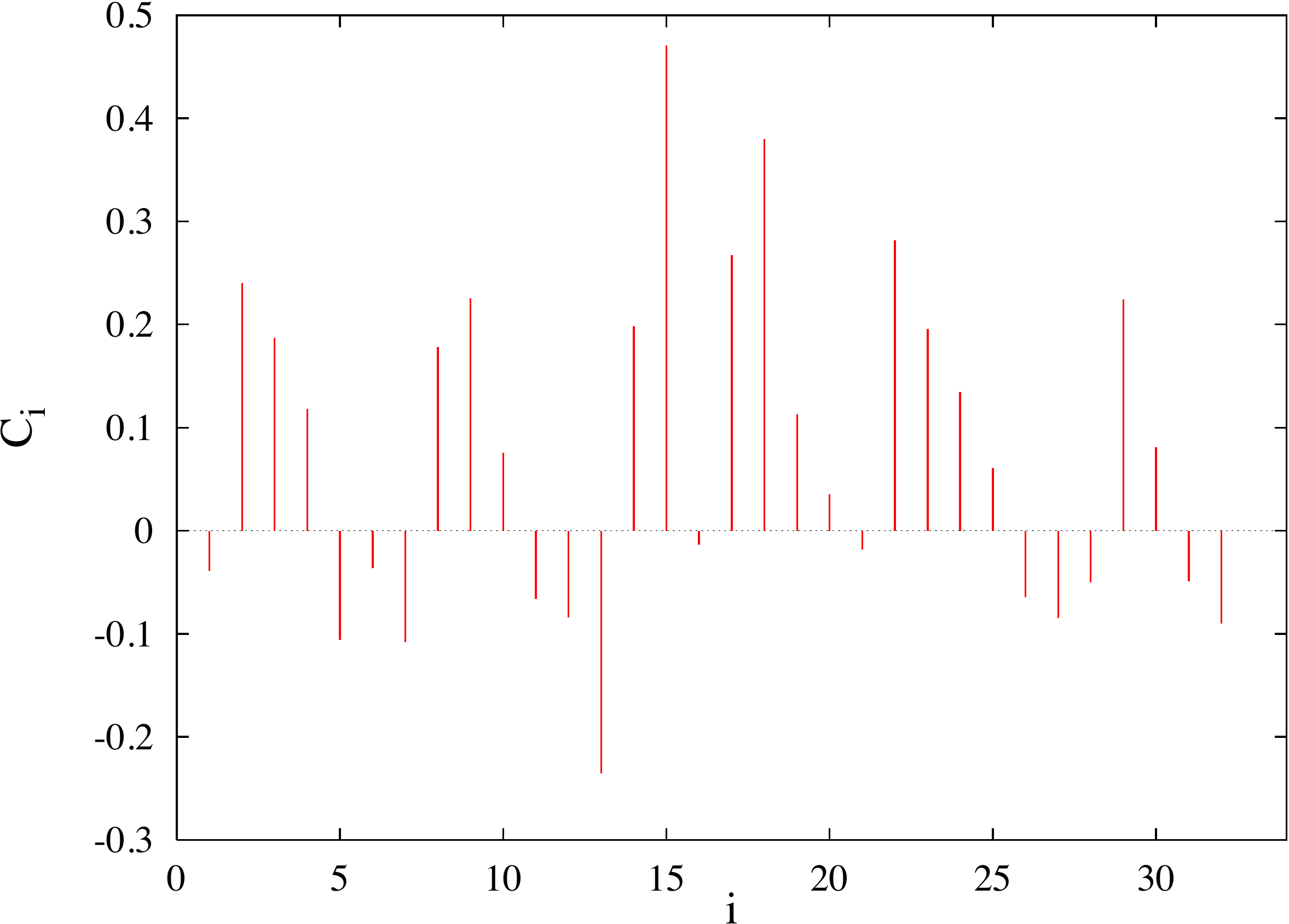}
\caption{(Color online)  Coefficients of the state of $N=5$ qubits
maximizing $QD$. No clear algebraic structure can be drawn from
these results. See text for details.} \label{fig3}
\end{center}
\end{figure}

\begin{figure}[htbp]
\begin{center}
\includegraphics[width=8.6cm]{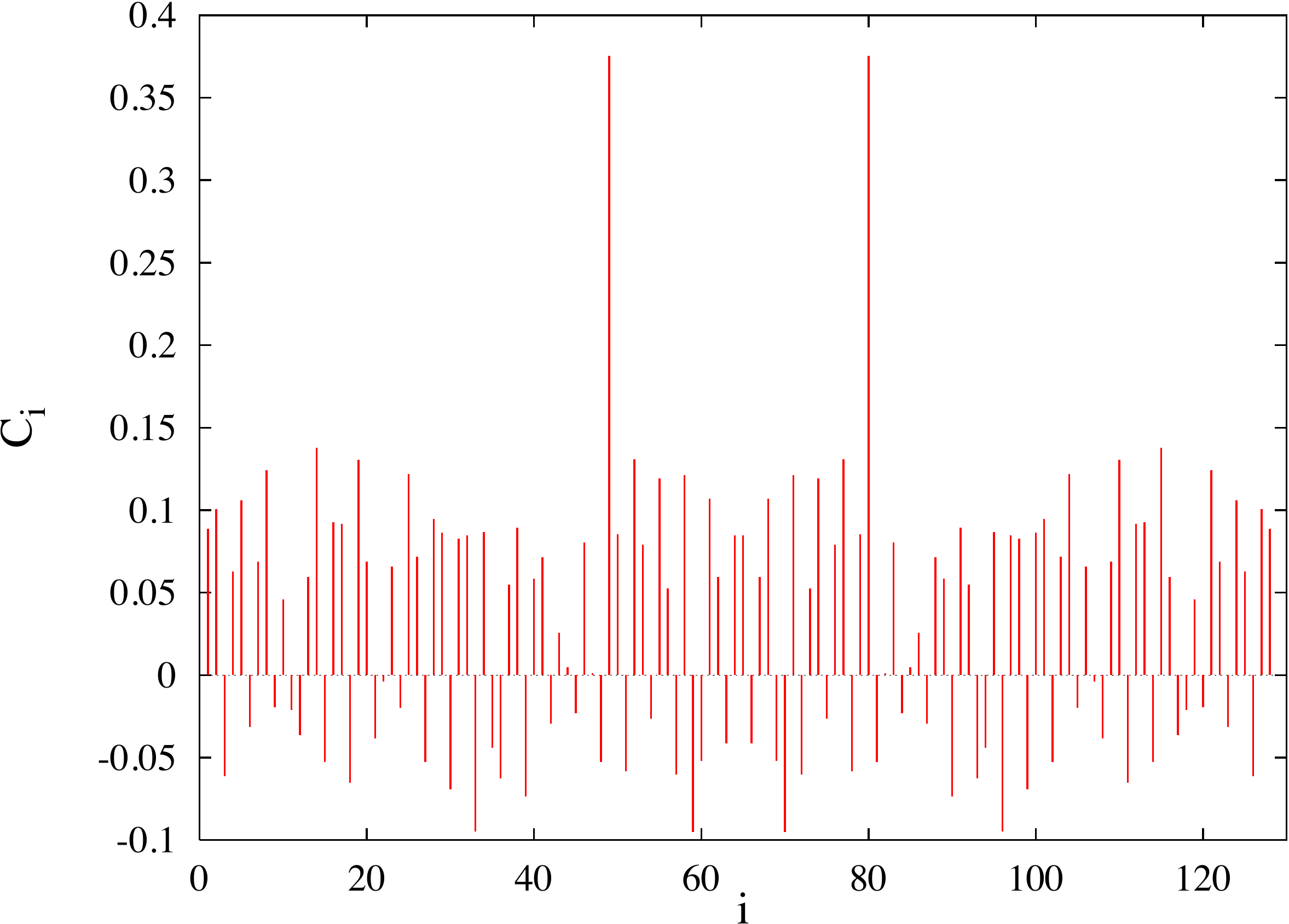}
\caption{(Color online) Similar plot as that of  Fig. 3 for $N=7$
qubits. A single  state ($|\Psi_{48,7}^{+} \rangle$) significantly
contributes to the total $QD$. Notice the symmetry between $i$ and
$2^N-i$ coefficients, which constitutes a clear sign of an
underlying superposition of maximally correlated states. See text
for details.} \label{fig4}
\end{center}
\end{figure}

\noindent The case of N=7 qubits case is tantalizing. As in the case
of N=5 qubits, we have not been able to find a simple algebraic
form. Per contra, our hypothesis of --as far as QD is concerned-- a
total symmetry in the state basis is duly confirmed here. That is,
the state of seven qubits is formed by a plethora ($2^{7}/2=64$) of
maximally correlated states. The state we obtain possesses a maximum
QD of 8.4290, which is only an approximate value. Fig. 4 shows the
value of the expansion-coefficient for each position in the
computational basis. The two peaks correspond to the state
$|\Psi_{48,7}^{+} \rangle$, which due to unknown reasons, becomes
differentiated from the rest. Once again, complex and real
coefficients are equivalent when providing states with maximum QD.

\begin{figure}[htbp]
\begin{center}
\includegraphics[width=8.6cm]{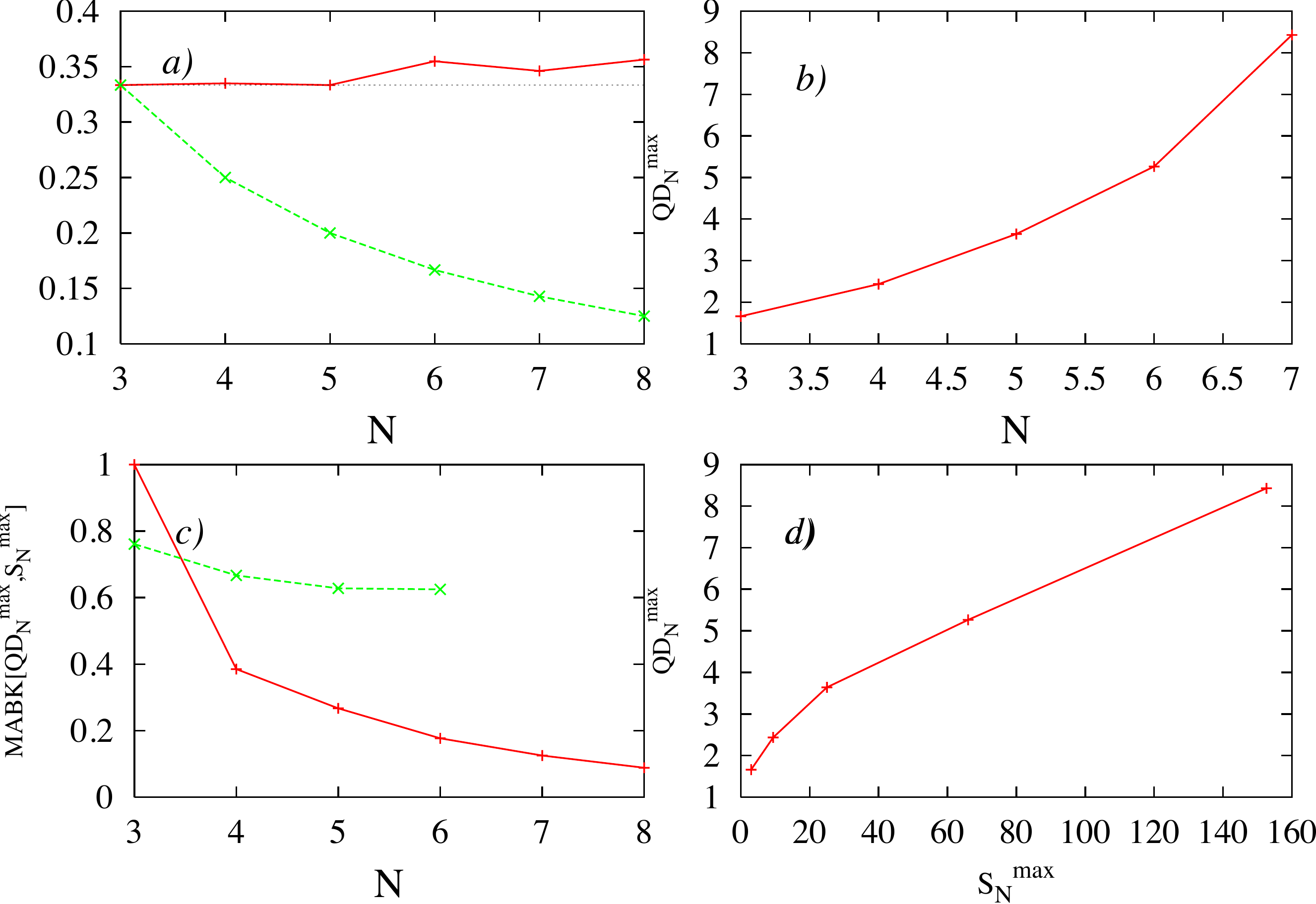}
\caption{(Color online) (a) Value of $S_N^{\max}$ vs $N$ in terms of
$[N (2^{N-1}-1)\frac{1}{3}\ln2]$ (upper curve). $N=3$, and $N=5$
coincide (1/3, horizontal line). The remaining values display a
step-like evolution. $S_N$ for $|GHZ_N\rangle$ states (maximally
nonlocal) in units of the same quantity $[N
(2^{N-1}-1)\frac{1}{3}\ln2]$ is also depicted (lower curve). (b)
Value of $QD_N^{\max}$ vs $N$. The evolution is clearly exponential.
(c) Normalized non-locality (to $2^{\frac{N+1}{2}}$) for maximally
entangled (solid line) and discordant (dot-dashed line) qubit
states. The $QD$ case saturates to $k2^{\frac{N+1}{2}}$, $k$ being
some positive constant, for an increasing number of parties. (d)
$QD_N^{\max}$ vs. $S_N^{\max}$. Notice that each point corresponds
to different states. On the whole, a monotonic increasing behavior
is apparent. See text for details.} \label{fig5}
\end{center}
\end{figure}

\section{Comparison between different measures}

\noindent At first sight, there is no precise way to correlate
measures $S, MABK_N^{\max}$, and $QD$. However, although  different
in nature, they globally behave in analogous fashion. We  observe
that, in all three cases, the states's maximal measures  for each
quantum correlation's type either increase or diminish exponentially
with the number of parties. This behavior becomes apparent in Fig.
5. Now, when relating two definite measures, special instances
appear. Let us consider the case of N=4 qubits. It was shown that
two states reached maximum ``real'' and ``complex'' entanglement
separately. Surprisingly, the more entangled the state was, the
lesser non-locality it possessed. Additionally, the state maximizing
entanglement is the only one in all multiqubit states here
considered that has non zero $QD$. In general, entanglement and
quantum discord are more similar to each other than to non-locality.

\noindent   Regarding non-locality vs. entanglement, it is plain
from Table II that maximum entanglement by no means implies
maximum non-locality. Nevertheless, $MABK_N^{\max}$ is not that
different from maximum entanglement. This fact is easily seen when
analyzing generalized GHZ states, which are the ones that maximize
MABK Bell inequalities. Measure $MABK_N^{\max}$ is equal to
$2^{\frac{N+1}{2}}$, and its entanglement is
$S_N(|GHZ_N\rangle)=(2^{N-1}-1)\ln2$ ($QD_N(|GHZ_N\rangle)=0$).
Entanglement and $MABK_N^{\max}$ for generalized GHZ states are
monotonic increasing functions one of the other $\rightarrow$
$S_N=(\frac{1}{4}[MABK_N^{\max}]^2-1)\ln2$. The average
entanglement per partition is equal to $\ln2$ in all cases.

\noindent   When computing the maximum entanglement for qubits
systems, we obtain that $S_N^{\max}\ge N
(2^{N-1}-1)\frac{1}{3}\ln2$, equality holding for $N=3,5$. In Fig. 5
a) we depict $S_N^{\max}$ and $S_N(|GHZ_N\rangle)$ in terms of $[N
(2^{N-1}-1)\frac{1}{3}\ln2]$ vs. $N$. What we obtain is that
entanglement for maximally entangled states  behaves in the same
way, differences appearing between clusters of states (states of
3,4, and 5 qubits are to be compared with results for 6,7, and 8
qubits). A kind of  step-like behavior is apparent, as well as the
decreasing tendency of $S_N(|GHZ_N\rangle)$. This fact constitute a
strong argument concluding that non-locality vs. entanglement
display a quite special behavior as opposed to the relation
entanglement vs. discord, where general tendencies are more
pronounced.

\noindent   $QD_N^{\max}$ for each number of qubits is depicted in
Fig. 5 b). This curve is perfectly fitted by an exponential. Thus,
maximum $QD$ also evolves exponentially with the number of parties
involved. As far as non-locality is concerned, Fig. 5 c) displays
the evolution of $MABK$ for those states with maximal discord
(dot-dashed line) and maximum entanglement (solid line) in units of
$2^{\frac{N+1}{2}}$. It is plain that non-locality for maximally
entangled states diminishes exponentially, whereas in the case of
quantum discord, a sort of proportionality is reached
($MABK(QD_N^{\max}) \propto 2^{\frac{N+1}{2}}$ for $N\ge 7$).

\noindent   When comparing maximum quantities corresponding to
different states, we obtain a result such as the one depicted in
Fig. 5 d). That is,  the QD for those states maximizing QD
monotonically increases with their entanglement counterpart.

\begin{table}[h]
\begin{center}
\begin{tabular}{|c||c|c|c|}
  \hline
  $N$ & $MABK(S_{max})$ & $MABK(QD_{max})$ & $S(QD_{max})$\\
  \hline
  $3$ & 4 - 100\% & 3.0430 - 76\% & $2.7394\ln2$ - 91\% \\
  $4$ & 2.1773 - 38\% & $\frac{8}{3}\sqrt{2}$ - 67\% & $8\ln2$ - 85\% \\
  $5$ & 2.1381 - 27\% & $\approx 5.0233$ - 63\% & $17.7864\ln2$ - 71\% \\
  $6$ & 2 - 18\% & $5\sqrt{2}$ - 62.5\% & $41.2542\ln2$ - 63\%  \\
  $7$ & 2 - 13\% & $\approx 7.2524$ - 45\% & $\approx 97.9457\ln2$ - 64\%  \\
  $8$ & 2 - 9\%  & - & - \\
  \hline
\end{tabular}
\end{center}
\caption{Several types of correlation measure for states with
increasing number of qubits. The first column displays de maximum
violation of the generalized MABK Bell inequality for those states
that maximize the generalized measure of entanglement, as well as
the concomitant percentage with respect to the maximum possible
violation ($2^{\frac{N+1}{2}}$ for each $N$). The second row
displays similar results for those states with maximum QD. Finally,
in the last column, the entanglement for states maximizing QD is
computed. Notice that some results are analytic. The computation of
QD for states with maximum entanglement is not depicted, as it
vanishes except for N=4 ($QD(S_{max})=0.7548 - 31\%$). See text for
details.} \label{2}
\end{table}

\section{Conclusions}

\noindent   We have explored the tripod
entanglement-discord-non-locality for multiqubit systems. We have
revisited the known cases of maximally entangled states of $N=$ 4-,
5-, and 6-qubits, exploring their properties as far as non-locality
is concerned. Specifically, the cases of 7 and 8 qubits have been
studied in some detail, and allowed for original results in both
cases. It has become clear that, some similarities notwithstanding,
entanglement and non-locality $MABK_N^{\max}$ behave differently in
one or the other direction. That is, the more entangled the states
are, the less nonlocal they become. Recall that, however, we only
focus on extreme cases, and not on arbitrary states.  The
aforementioned tendency has been confirmed for an increasing number
of qubits. This is the case because entanglement and non-locality
constitute, after all, different resources.

\noindent   Quantum discord, which has been recently proved to be a
valuable resource, has been extended here to multiqubit systems by
introducing a measure over all two-qubit partitions. Employing the
corresponding maximum value, we have discovered states of ``maximum
discord'' for up to 8 qubits. These states possess very interesting
features, including unexplained asymmetries among their parties.
Furthermore, we strongly believe that these maximal $QD-$states may
find interesting applications in quantum information processing.

\noindent   In the light of recent research, we have also explored
entanglement in multiqudit systems, and obtained the concomitant
maximally entangled states for different $(N,D)-$configurations.
Also, the way of defining quantum mechanics over real or complex
numbers is subject to inspection here. We find that  real
expansion-coefficients in a given basis  suffice to explain maximum
entanglement. Recall that this has not been the case for multiqubit
states.

\noindent   Finally, the comparison of all types of measure leads us
to conclude that entanglement and discord are both positively
correlated, whereas non-locality decreases with either maximum
entanglement or discord. We conclude, thus, that maximum
entanglement and maximum discord display  some similarities, whereas
this scenario is blurred when considering non-locality.

\appendix
\section{On correlations measures}

\nd  Research on the properties and applications of multipartite
entanglement measures has attracted considerable attention in
recent years
  \cite{BSSB05,HS,MW02,B03,WH05a,WH05b,CZ05,LS05,S04,CMB04,AM06,CHDB05}.
 One of the first useful entanglement measures for $N$-qubit pure
states $|\phi \rangle$ to be proposed was the one introduced by
Meyer and Wallach \cite{MW02}. It was
 later pointed out by Brennen \cite{B03}
 that the measure advanced by Meyer and Wallach
 is equivalent to the average of all the single-qubit
 linear entropies, that is, the average entanglement of each qubit
 of the system with the remaining $(N\!-\!1)$-qubits.

 Another way of characterizing the global amount of entanglement
exhibited by an $N$-qubit state is provided by the sum of the
(bi-partite) entanglement measures associated with the
 $2^{N-1}-1$ possible bi-partitions of the $N$-qubits system
\cite{BSSB05}. This particular number takes into account that the
marginal density matrices describing the $k$th party,
  after tracing out the rest,  are equivalent to those of $N-k$
  parties  because of the relation $\binom {N} {k}=\binom {N} {N-k}$. In essence,
 these entanglement measures are given by the degree of mixedness of the marginal density
matrices associated with each bi-partition. In our case, we shall
use the von Neumann entropy

\begin{equation} \label{Smesure}
S_{VN} = \sum_i-Tr[\rho_i \ln \rho_i] ,
\end{equation}

\noindent where the sum is performed over all $2^{N-1}-1$ different
bipartitions.

\nd A considerable amount of research has been devoted to unveil
the mathematical structures underlying entanglement, in particular
concerning those states which possess maximum entanglement, as
given by some appropriate measure. The survey of (pure) states
maximizing measure (\ref{Smesure}) has been the subject of intense
research, initiated by the work of Brown et al. \cite{BSSB05}.
These {\it genuinely entangled states} achieving the maximum
possible entanglement according to the (\ref{Smesure}) have
recently investigated with regards to their usefulness  for
perfect quantum teleportation, superdense coding, and quantum
secret sharing. Indeed, highly entangled multipartite states
generate intense interest for quantum information processing and
one-way universal quantum computing \cite{Briegel}. They are
essential for several quantum error codes and communication
protocols \cite{Cleve}, since they are robust against decoherence.
{\it The aforementioned research for maximally entangled states
will be extended here  to N=7 and N=8 qubits in Section III}.


\nd Quantum discord is also a valuable resource for the
implementation of non-classical information processing protocols
\cite{20,ferraro,dattaprl,luo,datta}. In the light of these
developments, it becomes imperative to characterize multipartite
states not only through their maximum amount of entanglement but
also through the maximization of their corresponding total quantum
discord, establishing  the links within the tripod
entanglement-discord-nonlocality.

\nd In addition to entanglement, quantum discord \cite{6,20}
constitutes a rather recent information-theoretical measure of the
``non-classicality" of bipartite correlations as given by the
discrepancy between the quantum counterparts of two classically
equivalent expressions for the mutual information. Quantum discord
corresponds to a new facet of the ``quantumness" that arises even
for non-entangled states. More precisely, quantum discord is defined
as the difference between two ways of expressing (quantum
mechanically) such an important entropic quantifier. Let $\rho$
represent a state of a bipartite quantum system consisting of two
subsystems $A$ and $B$. If $S(\rho)$ stands for the von Neumann
entropy of matrix $\rho$ and
 $\rho_A$ amd $\rho_B$ are the reduced (``marginal") density matrices
describing the two subsystems, the quantum mutual information (QMI)
$M_q$ reads \cite{6} \be \label{uno} M_q(\rho)= S(\rho_A) +
S(\rho_B) - S(\rho). \ee This quantity is to be compared to another
quantity ${\tilde M}_q(\rho)$, expressed using conditional
entropies, that classically coincides with the mutual information.
To define ${\tilde M}_q(\rho)$ we need first to consider the notion
of conditional entropy. If a complete projective measurement
$\Pi_j^B$ is performed on B and (i) $p_i$ stands for
$Tr_{AB}\,\Pi_i^B\,\rho$ and (ii) $\rho_{A|\vert \Pi_i^B}$ for
$[\Pi_i^B\,\rho\,\Pi_i^B/p_i]$, then the conditional entropy becomes

\be  \label{unobis}  S(A \vert\,\{ \Pi_j^B \}) =\sum_i\,p_i\,
S(\rho_{A|\vert \Pi_i^B}), \ee and ${\tilde M}_q(\rho)$ adopts the
appearance

\be \label{dos} {\tilde M}_q(\rho)_{\{ \Pi_j^B \}} = S(\rho_A)- S(A
\vert\,\{ \Pi_j^B \}). \ee Now, if we minimize over all possible $
\Pi_j^B$ the difference $M_q(\rho)-{\tilde M}_q(\rho)_{\{ \Pi_j^B
\}}$ we obtain the $QD$, that quantifies non-classical correlations
in a quantum system {\it  including those not captured by
entanglement}. The most general parameterization of the
corresponding local measurements that can be implemented on one
qubit (let us call it B) is of the form $\{ \Pi_B^{0^{\prime}}=I_A
\otimes |0^{\prime}\rangle \langle 0^{\prime}|,
\Pi_B^{1^{\prime}}=I_A \otimes |1^{\prime}\rangle \langle
1^{\prime}|\}$. More specifically we have

\begin{eqnarray} \label{unitarity}
 |0^{\prime}\rangle &\leftarrow& \cos\alpha' |0\rangle + e^{i\beta'}\sin\alpha'|1\rangle \cr
 |1^{\prime}\rangle &\leftarrow& e^{-i\beta'}\sin\alpha'|0\rangle - \cos\alpha' |1\rangle,
\end{eqnarray}
\noindent which is obviously a unitary transformation --rotation in
the Bloch sphere defined by angles $(\alpha',\beta')$-- for the B
basis $\{|0\rangle,|1\rangle\}$ in the range $\alpha' \in [0,\pi]$
and $\beta' \in [0,2\pi)$. The previous computation of the QD has to
be carried out numerically, unless the two qubit states belong to
the class of the so called X-states, where QD is analytic
\cite{Xstates}. Accordingly, we shall introduce here a generalized
QD measure for multiqubit states in the form

\begin{equation} \label{QDmeasure}
QD = \sum_i QD[\rho_i],
\end{equation}

 \noindent where the sum  takes place only over all $N(N-1)/2$ reduced two qubit states $\rho_i$, since
 QD is only defined in that case.

 \noindent Since the evaluation of QD (\ref{QDmeasure}) involves an optimization
procedure and analytical results are known only in a few cases, an
interesting alternative was advanced in \cite{20} by introducing a
geometric measure of quantum discord (GQD). Let $\chi$ be a generic
zero QD state. The GQD measure is then given by Hilbert-Schmidt norm

\be \label{tres} GQD(\rho)= {\rm Min}_{\chi}[||\rho-\chi||^2],  \ee

\noindent where the minimum is taken over the set of zero-discord
states $\chi$. Given the general form of an arbitrary two-qubits
state
 in the Bloch representation
\ben \label{rhoBloch}& 4\rho=   \mathcal{I} \otimes \mathcal{I} +
\sum_{u=1}^{3} x_u \sigma_u \otimes \mathcal{I} + \sum_{u=1}^{3} y_u
\mathcal{I} \otimes \sigma_i + \cr & +\sum_{u,v=1}^{3} T_{uv}
\sigma_u \otimes \sigma_v, \een \noindent with $x_u=Tr(\rho
(\sigma_u \otimes \mathcal{I}))$, $y_u=Tr(\rho (\mathcal{I} \otimes
\sigma_u))$, and $T_{uv}=Tr(\rho (\sigma_u \otimes \sigma_v))$, it
is found in Ref.  \cite{20} that a necessary and sufficient
criterion for witnessing non-zero quantum discord is given by the
rank of the correlation matrix

\begin{equation} \label{Rmatrix}
\frac{1}{4} \left( \begin{array}{cccc}
1 & y_1 & y_2 & y_3\\
x_1 & T_{11} & T_{12} & T_{13}\\
x_2 & T_{21} & T_{22} & T_{23}\\
x_3 & T_{31} & T_{32} & T_{33}
\end{array} \right),
\end{equation}

\noindent that is, a state $\rho$ of the form (\ref{rhoBloch})
exhibits finite quantum discord iff the matrix (\ref{Rmatrix}) has a
rank greater that two. It is seen that the  geometric measure
(\ref{tres}) is of the  form \cite{20}

\ben \label{Dfinal} & GQD(\rho)=\frac{1}{4} \bigg( ||{\bf x}||^2 +
|| T ||^2  - \lambda_{\max} \bigg)=\cr &=
\frac{1}{R}-\frac{1}{4}-\frac{1}{4}\bigg( ||{\bf y}||^2 +
\lambda_{\max} \bigg), \een \noindent where $||{\bf x}||^2=\sum_u
x_u^2$, $\lambda_{\max}$ is the maximum eigenvalue of the matrix
$(x_1,x_2,x_3)^t (x_1,x_2,x_3) + TT^t$ and $R=1/Tr(\rho^2)$.  The
study of the properties of the generalized GQD will be here carried
out for N=3 and N=4 qubits, its form being analogous to that of
(\ref{QDmeasure}).


\nd Entanglement, however, is not the only quantum correlation
that can be considered in multipartite quantum systems (qubits in
our case). Since the formalization by Werner \cite{W89} of the
modern concept of quantum entanglement it has become  clear that
there exist entangled states that comply with all Bell
inequalities (BI). This entails that non-locality, associated to
BI-violation, constitutes a non-classicality manifestation
exhibited only by just a subset of the full set of states endowed
with quantum correlations.

\nd In some cases, however, entangled states are useful to solve a
problem if and only if they violate a Bell inequality
\cite{comcomplex}. Moreover, there are important instances of
non-classical information tasks that are based directly upon
non-locality, with no explicit reference to the quantum mechanical
formalism or to the associated concept of entanglement
\cite{device}. Most of our knowledge on Bell inequalities and their
quantum mechanical violation is based on the CHSH inequality
\cite{CHSH}. The scenario with two dichotomic observables per party,
 is the simplest one \cite{Collins} endowed with a nontrivial Bell inequality for the
bipartite case (with binary inputs and outcomes). Quantum
mechanically, these observables reduce to ${\bf A_j}({\bf
B_j})=\bf{a_j}(\bf{b_j}) \cdot \bf{\sigma}$, where
$\bf{a_j}(\bf{b_j})$ are unit vectors in $\mathbb{R}^3$ and
$\bf{\sigma}=(\sigma_x,\sigma_y,\sigma_z)$ the Pauli matrices.
Violation of CHSH inequality requires the expectation value of the
operator $B_{CHSH}= {\bf A_1}\otimes {\bf B_1} + {\bf A_1}\otimes
{\bf B_2} + {\bf A_2}\otimes {\bf B_1}  - {\bf A_2}\otimes {\bf B_2}
$ to be greater than two. It is indeed an inequality that poses an
upper limit ($2\sqrt{2}$, the Tsirelson-bound \cite{Tsirelson}), to
quantum mechanical correlations between distant events. A relation
exists here with  the  discussion and experimental determination of
whether local variable model (LVM) variables are required for, or
even compatible with, the representation of experimental results. A
proper way to measure non-locality (for the two qubit state $\rho$)
uses  the quantity

\begin{equation} \label{resultat}
 B_{CHSH}^{\max} \equiv \max_{\bf{a_j},\bf{b_j}} Tr (\rho B_{CHSH}).
 \end{equation}

\noindent Similarly, non-locality in the three qubit case is given
by the violation of the Mermin's inequality \cite{MABK}. This
inequality was conceived originally in order to detect genuine
three-party quantum correlations, impossible to reproduce via LVMs.
The Mermin inequality states that  $Tr(\rho B_{Mermin}) \leq 2$,
where $B_{Mermin}$ is the Mermin operator

\begin{equation} \label{Mermin}
 B_{Mermin}=B_{a_{1}a_{2}a_{3}} - B_{a_{1}b_{2}b_{3}} - B_{b_{1}a_{2}b_{3}} - B_{b_{1}b_{2}a_{3}},
\end{equation}

\noindent where $B_{uvw} \equiv {\bf u} \cdot {\bf \sigma} \otimes
{\bf v} \cdot {\bf \sigma} \otimes {\bf w} \cdot {\bf \sigma}$ with
${\bf \sigma}=(\sigma_x,\sigma_y,\sigma_z)$ being the usual Pauli
matrices, and ${\bf a_j}$ and ${\bf b_j}$ unit vectors in
$\mathbb{R}^3$. Notice that the Mermin inequality is maximally
violated by Greenberger-Horne-Zeilinger (GHZ) states. As in the
bipartite case, we shall define the following quantity

\begin{equation} \label{MerminMax}
Mermin^{\max} \equiv \max_{\bf{a_j},\bf{b_j}}\,\,Tr (\rho
B_{Mermin})
\end{equation}

\noindent as a measure of the non-locality of the state $\rho$.
While in the bipartite case the CHSH inequality is the strongest
possible one, no bond equivalent to Tsirelson's is available for
arbitrary dimensions.

\noindent Now, in the case of multiqubit systems, one must instead
use a generalization of the CHSH inequality to N qubits. This is
done in natural fashion by considering an extension of the CHSH or
Mermim inequality to the multipartite case. The first Bell
inequality (BI) for four qubits was derived by Mermin, Ardehali,
Belinskii, and Klyshko \cite{MABK}. One deals with four parties with
two dichotomic outcomes each, the BI being maximum for the
generalized GHZ state $(|0000\rangle + |1111\rangle)/\sqrt{2}$. The
Mermin-Ardehali-Belinskii-Klyshko (MABK) inequalities are of such
nature  that they constitute extensions of older inequalities, with
the requirement that generalized GHZ states must maximally violate
them. To concoct an  extension to the multipartite case, we shall
introduce a recursive relation that will allow for more parties.
This is easily done by considering the operator

\begin{equation}
 B_{N+1}  \propto [(B_1+B_1^{\prime}) \otimes B_N + (B_1-B_1^{\prime}) \otimes B_N^{\prime}] ,
\end{equation}

 \noindent with $B_N$ being the Bell operator for N parties and $B_1={\bf v} \cdot {\bf \sigma}$,
 with ${\bf \sigma}=(\sigma_x,\sigma_y,\sigma_z)$ and ${\bf v}$ a real unit vector. The prime on the operator
 denotes the same expression but with all vectors exchanged. The concomitant maximum value

\begin{equation} \label{MABK_Nmax}
MABK_N^{\max} \equiv \max_{ \bf{a_j},\bf{b_j} }\,\,Tr (\rho {B_{N}})
\end{equation}

\noindent will serve as a measure for the non-locality content of a
given state $\rho$ of N qubits if ${\bf a_j}$ and ${\bf b_j}$ are
unit vectors in $\mathbb{R}^3$. The non-locality measure
(\ref{MABK_Nmax}) will be maximized by generalized GHZ states,
$2^{\frac{N+1}{2}}$ being the corresponding maximum value.

\noindent The MABK inequalities are not the only existing Bell
inequalities for N qubits \cite{MABKnew}, but they constitute a
simple generalization of the CHSH one to the multipartite case.
Accordingly, it will suffice to use these particular inequalities to
illustrate the basic results of the present work, as far as
non-locality is concerned.

 \nd Regardless of the quantity one shall compute or of the state one aims at finding, some sort
of optimization procedure must be carried out. In the case of
maximum entanglement (\ref{Smesure}), the $2^N$ coefficients of the
multipartite pure state constitute the variables to deal with (twice
if the state is of complex instead of real nature). For the quantum
discord QD, a minimization takes place for the two parameters in
every $N(N-1)/2$ reduced two qubit state, whereas non-locality
$MABK_{N}^{\max}$ requires an exploration among their corresponding
unit vectors defining the observers' settings. In any case, we have
performed a two-fold search employing i) an amoeba optimization
procedure, where the optimal value is obtained at the risk of
falling into a local minimum and ii) the well known simulated
annealing approach \cite{kirkpatrick83}. The advantage of this
computational 'duplicity' is that we can be confident regarding the
final results reached. Indeed, the second recipe contains a
mechanism that allows for  local searches that eventually can escape
 local optima.

\section*{Acknowledgements}

\nd J. Batle acknowledges fruitful discussions with J.
Rossell\'{o} and M. del M. Batle, while  A. Plastino and  M. Casas
acknowledge partial support under project FIS2011-23526 (MINECO)
and FEDER(EU).

\end{document}